\documentclass[preprint,showpacs,preprintnumbers,amsmath,amssymb]{revtex4}

\usepackage{graphicx}
\usepackage{dcolumn}
\usepackage{bm}

\usepackage{color}

\begin{document}

\preprint{APS/SandSwimmer}

\title{Swimming in Granular Media}

\author{Takashi Shimada$^{1,2}$}\email{shimada@ap.t.u-tokyo.ac.jp}
\author{Dirk Kadau$^{2}$}
\author{Troy Shinbrot$^{3}$}
\author{Hans J. Herrmann$^{2}$}
\affiliation{
${}^1$
Department of Applied Physics, Graduate School of Engineering, The University of Tokyo\\
7-3-1 Hongo, Bunkyo-ku, Tokyo, 113-8656 Japan,
}
\affiliation{
${}^2$
Computational Physics for Engineering Materials,
Institute for building materials, ETH Z\"urich,
}
\affiliation{
${}^3$
Department of Biomedical Engineering, Rutgers University
}
\date{\today}

\begin{abstract}
We study a simple model of periodic contraction and extension of large intruders in a granular bed
to understand the mechanism for swimming in an otherwise solid media.
Using an event-driven simulation, we find optimal conditions
that idealized swimmers must use to critically fluidize a sand bed
so that it is rigid enough to support a load when needed,
but fluid enough to permit motion with minimal resistance.
Swimmers - or other intruders - that agitate the bed too rapidly 
produce large voids that prevent traction from being achieved,
while swimmers that move too slowly cannot travel before the bed re-solidifies 
around them i.e., the swimmers locally probe the fundamental time-scale in a granular packing.
\end{abstract}


\maketitle

\begin{figure}[b]
\rotatebox{0}{\includegraphics[width=6.0cm]{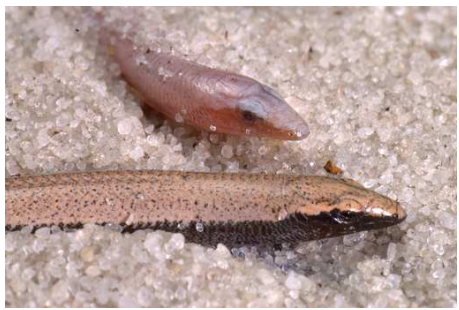}}
\caption{
A snapshot of sand swimmers ({\it Plestiodon reynoldsi}) emerging from a granular bed.
Figure credit: Alessandro Catenazzi.
}
\label{fig_lizard}
\end{figure}

Granular materials are able to act either as a solid,
supporting a load like sand on a beach, or as a fluid, e.g. flowing in an hourglass.
Some lizards are known to make use of the balance between 
solid-like and fluid-like properties to swim 
through sand beds (Fig. \ref{fig_lizard}).
Since they cannot survive on the surface of their desert habitats during the day,
they have developed the ability to submerge up to 10 cm beneath the sand
to lower temperature regions, and even to travel within the sand bed.
These reptiles exhibit a unique ability to manipulate the properties of granular materials, 
and their study stands to provide insight into the behavior of intruders in agitated granular 
beds more broadly,
a topic of ongoing investigation by research groups in diverse 
disciplines ranging from Physics\cite{phys1, phys2},
to Geology\cite{geo1, geo2},
and to Chemical Engineering\cite{chem1, chem2}.

There are at least two, and possibly more, mechanisms thought to be at work:
first, the swimmer can expand its hindquarters and thrust its
wedge-shaped head into the bed \cite{stebbins1944},
and second, it can submerge in an undulatory motion \cite{jayne2000}.
Neither of these behaviors is well understood,
and both appear to represent active manipulation of the 
properties of the bed to make it solid enough to 
provide traction, but fluid enough to minimize 
friction - indeed, recent investigations of the 
motion of robots on sand beds hinge on 
understanding this balance between solid-like and 
fluid-like properties of granular beds 
\cite{goldman2009}. 

\begin{figure}[t]
\rotatebox{90}{\includegraphics[width=6cm]{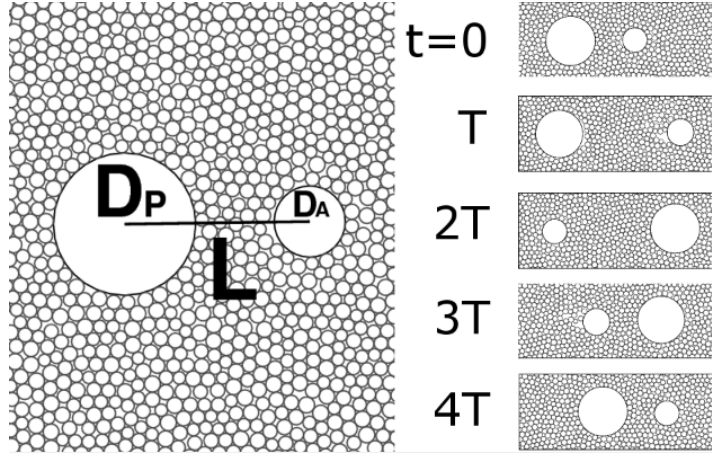}}
\caption{
Geometry of the swimmer (left)
and its shapes at the end of each cycle of the swimming stroke (right).
Movies of the swimming motion appear at
http://bopper.t.u-tokyo.ac.jp/\texttt{\symbol{126}}shimada/sandswim/movies.
}
\label{fig_snapshots}
\end{figure}

To start with a simple model, we have performed simulations based on a
recently proposed ``Pushme-Pullyou'' mechanism
\cite{avron2005}.
As shown in snapshots (Fig. \ref{fig_snapshots}, right),
this mechanism consists of contracting and protruding the
forequarters while enlarging the hindquarters (snapshots at $t = 0, T$),
then enlarging the forequarters to anchor the new position and
contracting the hindquarters ($t = 2T, 3T$),
and finally returning to the initial state ($t = 4T, 0$).
We find that this mechanism does produce steady
forward motion through a granular bed.
This mechanism differs from the well-known Purcell's swimmer
which functions in viscous fluids
\cite{purcell1977, shapere1987, wiggins1998, najafi2004, alexander2008},
however it is amenable to direct simulation,
provides insight into mechanisms that may be at work in granular systems [cf. SA Koehler et al., preprint 2008],
and may enable intriguing future technologies such as reconnaissance or
exploration for landmines beneath the desert floor.

To simulate this system, we define a two dimensional swimmer
consisting of anterior and posterior disks with diameters $D_A$ and $D_P$,
connected by a linear spring with spring constant $K$ and natural length $L_0 (t)$
(see Fig. \ref{fig_snapshots}, left).
Both the swimmer disks and the surrounding granular particles have hard core potentials.
The particle dynamics are modeled using an event-driven simulation method as described in
\cite{alder1962, lubachevsky1991, isobe1999}.
To make a swimming stroke,
we let the swimmer disks inflate and shrink,
and correspondingly we define the natural length of the connection between neighboring disks
to extend and retract.
We consider reciprocal swimming strokes
consisting of four cycles (A, B, C, and D, as illustrated in Fig. \ref{fig_snapshots} right)
with equal period ($T$) as:
\begin{equation}
(\dot{L_0}, \dot{D}_A, \dot{D}_P) =
\begin{cases}
	( \frac{L_0}{T},\ 0,\ 0 ) & (\mbox{A}: 0 \le t' < T) \cr
	( 0,\ \frac{\Delta D}{T},\ -\frac{\Delta D}{T} ) & (\mbox{B}: T \le t' < 2T) \cr
	( -\frac{L_0}{T},\ 0,\ 0) & (\mbox{C}: 2T \le t' < 3T) \cr
	( 0,\ -\frac{\Delta D}{T},\ \frac{\Delta D}{T}) & (\mbox{D}: 3T \le t') \cr
\end{cases},
\end{equation}
where $t' = ( t \mod 4T )$.
The initial shape of the swimmer is given as $(L_0(0), D_A(0), D_P(0)) = (L_0^-, D^-, D^+)$.
$\Delta D = D^+ - D^-$ 
is the difference between the maximum diameter $D^+$
and the minimum diameter $D^-$ of the swimmer disks.
$\Delta L_0 = L_0^+ - L_0^-$ is the difference between the maximum length $L_0^+$
and the minimum length $L_0^-$.
We set the bond to be very rigid (the spring constant $K \sim 10^4$)
so that the actual length $L(t)$ tightly follows $L_0(t)$ in the sand media.
In our work here, we fix the proportions of the swimmer to be
$L_0^+ / L_0^- = D^+ / D^- = 2$ and $L_0^- = 2.5 D^-$,
hence the swimming motion is specified by
the maximum diameter $D^+$ and the stroke time $4T$.
Note that the scaled swimming frequency
$F = D^+ / 4T$
gives the characteristic velocity of the swimming motion. 
We set the density of the swimmer at $\rho_{\mbox{swimmer}} = 0.68$,
so that the swimmer has approximately neutral density
with respect to the surrounding sand with volume fraction
$0.68$ (loosely packed case such as rapid swimming under free-boundary)
to $0.82$ (system with compacting top wall)
and its particle density $\rho_{\mbox{sand}} = 1.0$.

The sand particles are polydisperse,
defined to have diameters uniformly distributed on $1.0 \pm 0.15$
and we set the restitution coefficient to be $\epsilon_{\mbox{sand}} = 0.7$.
Collisions between sand particles and the swimmer are perfectly elastic ($\epsilon_{\mbox{swimmer}} = 1.0$).
To avoid inelastic collapse \cite{mcnamara1994},
we apply the TC model \cite{tc_model}:
if a particle has a collision in a very small time window
after its last collision, the collision is treated as elastic collision ($\epsilon_{\mbox{sand}} = 1$).
In this study we set $t_c = 10^{-3}$.
We normalize the simulations by setting gravity, $g$, to unity,
and we define the bottom of the system to be a fixed hard wall with specular reflection.
The left and the right boundaries are periodic.
For the top,
we have studied two different conditions:
a free boundary and a wall with constant pressure.
In the latter case,
we put a rigid and horizontal wall with specular reflection
on the top with line density $\rho_{\mbox{wall}} = 100$.
We use this condition to mimic swimming in the ``bulk'', i.e. far beneath the surface.

\begin{figure}[t]
\rotatebox{-90}{\includegraphics[width=6cm]{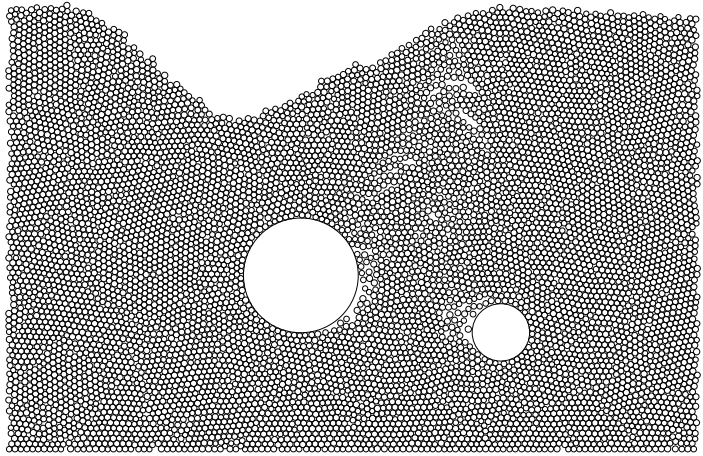}}
\caption{
A typical snapshot of the system with free boundary on top.
Note that as the swimmer moves,
it produces void spaces that propagate upward and modify the free surface (see text).
}
\label{freeboundary}
\end{figure}

When we have a free-boundary on the top,
dips are formed on the sand surface by the Pushme-Pullyou swimming strokes (Fig. \ref{freeboundary}).
In this situation, the swimming direction is unstable because of the interaction
between the swimmer and the free surface shape.
The voids produced by the swimmer modify the free surface
and change the stress field around the swimmer in a complex way.
Therefore we focus on swimming in the more straightforward ``bulk'' condition.
It is worth mentioning that we can modulate the swimming direction
by adding quick deflation and moderate inflation motion.
The surrounding sand cannot catch up with fast motion
therefore the disk drops to the bottom of the void shaped by the disk itself before the deflation.
In contrast, the disk almost keeps its position during the slower inflation cycle.
Hence the disk get downward shift in total.
However, a better control of direction is beyond the scope of this study.

\begin{figure}[t]
\rotatebox{-90}{\includegraphics[width=6cm]{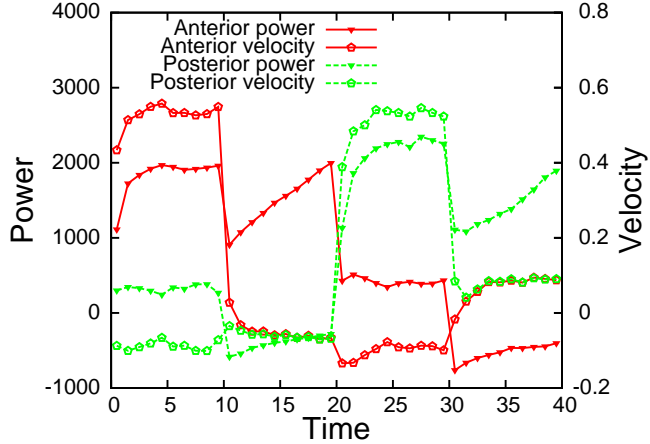}}
\caption{
Time evolutions of the power and the velocity of the disks in the bulk condition.
During cycle A,
the smaller anterior disk is extended;
during cycle B, the anterior disk expands and the posterior one contracts,
during cycle C the posterior disk is retracted,
and during cycle D the anterior disk contracts and the posterior one expands.
The parameters for this example are $D^+ = 20$ and $T = 40$ ($F=0.125$).
}
\label{profile}
\end{figure}

Providing a heavy top surface stabilizes the swimming direction.
So we can evaluate swimming speeds and efficiencies as a function of the model parameters.
In Fig. \ref{profile},
we show the time evolution of velocity and power expended by the disks
with moderate swimming frequency ($F = 0.125$),
in which only small voids are observed around the swimmer.
From the time evolutions shown, the swimming motion can be understood as follows.
The small velocity of the larger disk in cycle A (posterior disk) and cycle C (anterior disk)
implies that the larger disk is effectively anchored and moves little as the smaller disk extends and retracts.
We have confirmed that the drag force is independent of velocity
by simulations at multiple different frequencies.
This is characteristic for slow motion of large intruders in granular media
\cite{drag_granular}.

In contrast, during the inflating cycle
(cycle B for the anterior disk and cycle D for the posterior disk)
the inflating disk is held at a constant pressure $p$,
i.e., the sand around the swimmer appears to become fluidized
and consequently the power that the inflating disk exerts grows proportionally to time.
This picture changes during the shrinking cycle at the same frequencies:
the relaxation of the surrounding sand particles cannot catch up with the swimmer's motion,
hence the swimmer cannot retrieve the energy it has expended.
From this observation,
one can roughly estimate the total displacement $\Delta x$
and the total work $W$ done by the swimmer during one cycle to obey:
\begin{eqnarray}
\Delta x &\sim& \Delta L_0 \propto D^+, \\
W &\sim& \frac{\pi p}{4} \left\{ (D^+)^2 - (D^-)^2 \right\} + 2 \eta D^- \Delta L_0 \propto S,
\label{eqWork}
\end{eqnarray}
where $p$ and $\eta$ are phenomenological pressure and dynamic drag coefficient,
and $S = \pi (D^+)^2 / 4$ is the maximum area of the swimming disk.

\begin{figure}[t]
\rotatebox{-90}{\includegraphics[width= 6cm]{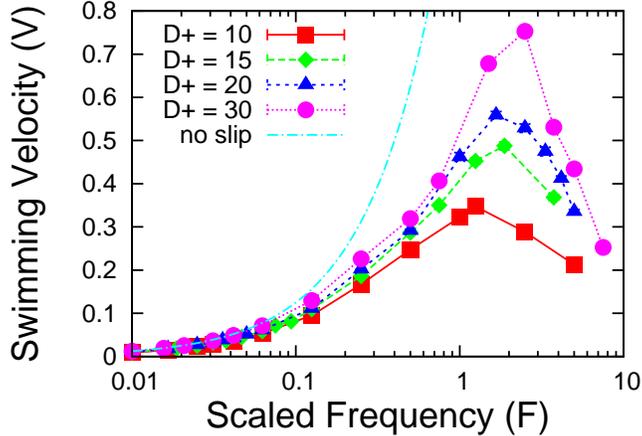}}
\caption{
Plot of the swimming velocities.
The line ``no slip'' corresponds to the velocity 
expected for the swimming without slip: $V = 1.25F$.
}
\label{fig_velocity}
\end{figure}

Let us now focus on two important quantities of the motion:
the swimming velocity
$V = \Delta x / 4T$
and the swimming efficiency
$E = S \Delta x / W$.
The velocity and the efficiency as functions of swimmer size and frequency
are shown in Fig. \ref{fig_velocity}.
In the slow-motion region,
the swimming velocity increases linearly with the swimming frequency $F$
since the total displacement obtained by each stroke is constant.
As the swimming frequency is further increased,
the resulting swimming velocity reaches a maximum and finally decreases.
The decrease in swimming speed coincides with an enlargement in void space surrounding the swimmer:
it appears that when the agitation is sufficiently rapid,
a low density bubble forms around the swimmer,
preventing it from gaining traction and moving forward.
As a result, there is an optimum frequency for swimming velocity.

\begin{figure}[t]
\rotatebox{-90}{\includegraphics[width=6cm]{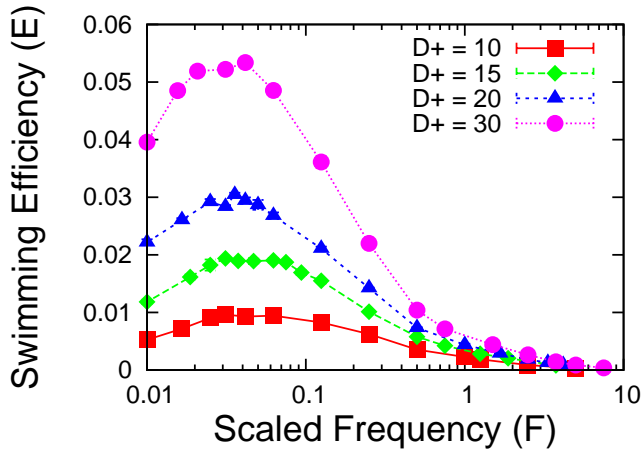}}
\caption{
Plot of the swimming efficiencies.
All swimmers have optimum swimming frequencies.
}
\label{fig_efficiency}
\end{figure}
\begin{figure}[t]
\rotatebox{-90}{\includegraphics[width=6cm]{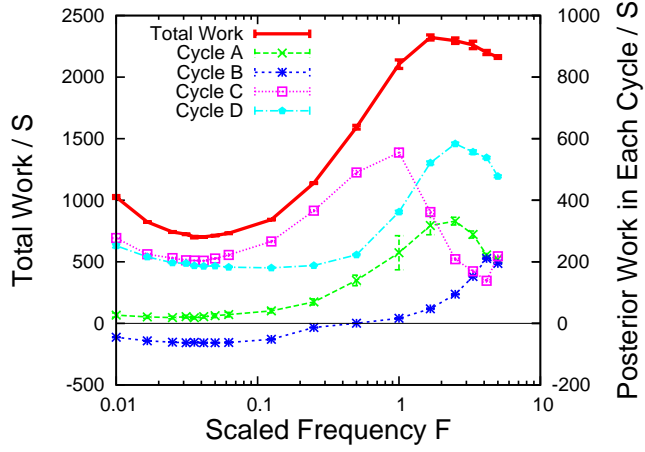}}
\caption{
Total work done by the swimmer with diameter $D+ = 20$ scaled by $S$ (bold line)
and the corresponding scaled works done in each cycle by the posterior disk (thin lines).
Note that the posterior disk anchors, deflates, is dragged, and inflates
in cycle A, B, C, and D, respectively.
}
\label{fig_work}
\end{figure}

What is more striking is that
there is a second, different, swimming frequency at which the swimming efficiency is maximal
(Fig. \ref{fig_efficiency}).
The swimming efficiency takes its maximum in the slow-swimming regime,
in which the swimmer does not slip against the bed.
This implies that the efficiency maximum comes from the minimum in the total work $W$.
In Fig. \ref{fig_work}, we confirm this:
the total work shows a peak near the point at which
the peak of the swimming velocity appears (scaled frequency $F \sim 2$),
after which the stroke starts slipping.
In the low frequency limit,
we expect the total work expended over a swimming cycle to approach a finite value,
unlike the case of swimming in fluid, where the work approaches zero.
Our simulation results indeed suggest this non-zero quasi-static limit.
In addition, it is found that the total work $W$
takes a minimum at certain low, non-zero frequencies.
How and why does the work exhibit a minimum before the quasi-static limit?
Looking at the work done by one disk in each cycle separately,
we find that the dragging cycle (cycle A for the anterior disk and cycle C for the posterior disk)
and the inflating cycle (cycle B for the anterior disk and cycle D for the posterior disk)
contribute most to the minimum (Fig. \ref{fig_work}).
In contrast,
the work done by the anchoring disk
(cycle C and A for the anterior and the posterior disks, respectively)
and the work retrieved to the deflating disk 
(cycle D and B for the anterior and the posterior disks, respectively)
converge to a static limit around $F = 0.1$.

This can be understood in more detail by scrutiny of the power expended by the swimmer.
As displayed in Fig. \ref{profile} and argued by Eq. (\ref{eqWork}),
we find the linearly increasing form in the inflating cycle 
under conditions that the disk feels constant pressure from the surrounding sands.
As frequency decreases,
the profile shifts to a rectangular shape which is similar to the dragging cycle.
The profile in the dragging cycle keeps its rectangular shape for lower frequency,
while increasing its height.
Since the temporal evolution of the 
position of the wall on top is independent of the swimming frequency in the slowly swimming region,
the global density of the sand is considered to be constant.
Thus, the increase of the drag experienced by the disk as it deforms the surrounding sand
comes from the re-organization of force chains within the granular bed.
In other words, our simulations indicate that
sand swimming relies on simultaneous solidification (near the anchored disk)
and fluidization (near the moving disk) of the granular bed.
This in turn relies on the swimmer's ability to exploit
multiple timescales by moving one disk rapidly while the bed remains slightly fluidized,
and holding the other disk static in its solidified environment.
The swimmer at the efficiency optimum takes advantage of this difference in timescales
to minimize its energy expenditure.
We speculate that more realistic models of sand swimming may
depend on a similar balance between timescales in the granular bed.

In conclusion,
we have studied swimming in sand under Avron's Pushme-Pullyou mechanism \cite{avron2005}
using an event-driven simulation.
We find that
both swimming velocity and swimming efficiency have optima at different swimming frequencies.
If the swimmer moves too fast,
the voids created by the swimming stroke
become larger and finally surround the whole swimmer.
In this situation the swimmer slips and
therefore the swimming velocity exhibits its maximum in fast motion region.
If the swimmer moves too slowly, the 
bed re-solidifies before the swimmer can move 
forward.  Optimal efficiency is achieved when the 
swimmer achieves a balance between these 
extremes: the inflating and moving disk is 
surrounded by weakly fluidized grains, while the 
anchoring disk is surrounded by solidified sand.  
These results suggest that
the interplay between coexisting
fluidized, slightly fluidized, and static granular states may be important for the
understanding more complex situations including swimming near a free surface,
the interaction between several swimmers
\cite{alexander2008, ortiz2005, ishikawa2008},
and more realistic shapes of swimmers.
The frequency and size of the swimmer, i.e. its characteristic parameters,
determine the time-scale over which the fluidization-solidification transition can be induced.
In that sense our swimmer constitutes a simple device
to locally probe this fundamental time-scale in a granular packing.

\begin{acknowledgments}
The authors thank S. Luding and S. Koehler for crucial advice,
and A. Catenazzi for providing photographs of sand swimmers.
This work was initiated during the visit of T. Shimada to ETH,
which is supported by Grants to Fund Long-term Visits to Overseas Research Institutes
by the University of Tokyo Academic Staff.
\end{acknowledgments}

\bibliography{condmatSandSwim.bbl}
\end{document}